# Micro-Brillouin Study of the Eigenvibrations of Single Isolated Polymer Nanospheres

Y. Li, H. S. Lim, Z. K. Wang, S. C. Ng, and M. H. Kuok[*]
*Department of Physics, National University of Singapore,
2 Science Drive 3, Singapore 117542, Singapore*

The localized acoustic modes of single isolated polymethyl methacrylate (PMMA) and polystyrene nanospheres have been studied by micro-Brillouin light scattering. The measured mode frequencies are analyzed on the basis of the Lamb theory formulated for a sphere under free boundary conditions. By measuring light scattering from single isolated particles, placed atop a piece of polished silicon wafer, the free-surface conditions are almost experimentally realized. The observed spectral peaks are attributed to localized eigenvibrations whose frequencies scale as inverse sphere diameter, in accordance with Lamb's theory. The Young's moduli and Poisson ratios of the polymer spheres studied have been evaluated from fits to the experimental data. We have demonstrated that micro-Brillouin spectroscopy is a powerful technique for probing the acoustic dynamics and mechanical properties of nanostructures.

Corresponding Author :

Meng Hau Kuok
Department of Physics
National University of Singapore
2 Science Drive 3
Singapore 117542
SINGAPORE
Telephone:     + (65) 65162609
Fax:           + (65) 67776126
Email:         phykmh@nus.edu.sg

---

[*] Author to whom correspondence should be addressed.

# 1. INTRODUCTION

An understanding of the nature of acoustic phonons of nanostructured materials is important because they are closely related to the mechanical and thermal properties of these materials. Additionally, a knowledge of the mechanical and thermal properties of nanostructures is required for their use as structural and functional elements in devices. Recently, photonic glasses composed of randomly-arranged polymer nanospheres have been synthesized [1]. Photonic glasses are interesting as they, despite their short-range order, have been discovered to have optical properties qualitatively similar to those of photonic crystals [2].

Brillouin light scattering is particularly suited to the investigation of acoustic modes of nanostructures [3-6]. Besides being a non-contact and non-invasive optical technique, its attributes include the ability to measure the acoustic dynamics of nano-objects of any shape. In this study, the acoustic dynamics of polymethyl methacrylate (PMMA) and polystyrene nanospheres have been investigated by micro-Brillouin spectroscopy. The motivation for studying these polymer nanospheres is their applications in areas such as photonic crystals [7] and, drug and gene delivery [8]. Brillouin data on these polymer spheres were interpreted on the basis of the Lamb theory [9] formulated for a sphere under free surface conditions. The eigenvibrations of *single isolated* nanospheres, placed atop the polished surface of a piece of silicon wafer, were studied by inelastic light scattering. These experimental conditions are a near realization of the ideal free-surface ones specified in the Lamb theory. The recording of spectra of single isolated nanospheres was achieved using a micro-Brillouin arrangement comprising a high-resolution microscope optically coupled to a conventional Brillouin light scattering system.

According to the Lamb theory, the acoustic eigenvibrations of a sphere are classified into spheroidal and torsional modes, with angular momentum quantum number $l = 0, 1, 2, \ldots$, and $l = 1, 2, 3, \ldots$ respectively. The sequence of vibrational eigenmodes, labeled by $(n, l)$, is indexed by $n$ $(= 1, 2, 3, \ldots)$ in increasing order of energy. Selection rules for inelastic light scattering permit the observation of only spheroidal modes [10].

## 2. EXPERIMENTAL DETAILS

The PMMA spherical samples used (Bangs Laboratories, Inc) have respective diameters (*d*) of 300 and 900 nm, while for the polystyrene ones (Polysciences, Inc) $d$ = 343, 400, 445, 494 and 597 nm. A typical SEM image, that of the 400 nm polystyrene spheres, is depicted in Fig. 1. Brillouin measurements were performed in the 180°-backscattering geometry with a modified DM/LM Leica microscope optically interfaced to a 6-pass Fabry-Perot interferometer. The spectra were excited with several milliwatts of the 514.5 nm radiation of an argon-ion laser. Experiments were conducted in the following procedure. An aggregate of loose monodisperse nanospheres were placed on the surface of a piece of polished silicon wafer mounted on a tiny rotation stage attached to the microscope's x-y translation stage. The microscope 100X objective lens served as both the focusing and the collection lens. A CCD camera attached to the microscope allowed viewing of the nanospheres and by moving the translation stage, a single isolated nanosphere could be selected. After it was brought into focus and irradiated with the laser light, spectral data acquisition could then commence. The typical scanning duration per spectrum is 2 hours.

## 3. RESULTS AND DISCUSSION

Representative micro-Brillouin spectra of a single isolated 494 nm polystyrene sphere and a single isolated 300 nm PMMA sphere are shown in Figs. 2 (a) and 2 (b) respectively. Each spectrum, in general, features well-separated sharp peaks. Figure 3 illustrates the difference between the micro-Brillouin spectrum of an aggregate of loose 300 nm PMMA spheres (4% size distribution) and that of one of its isolated spheres. Both spectra were recorded under the same experimental conditions. The spectral peaks of the aggregate are noticeably broadened relative to those of one of its component isolated spheres. Such linewidth broadening has been accounted for by the size polydispersity of the particles in the aggregate [5].

The experimental data on single isolated polymer spheres were analyzed on the basis of Lamb's theory [9]. Discussions will be confined to spheroidal modes, as torsional ones are not observable for inelastic light scattering. The frequencies $v_{nl}$ of spheroidal modes (*n,l*) can be calculated from the following equations.

$$2\left\{\eta^2 + (l-1)(l+2)\left[\frac{\eta j_{l+1}(\eta)}{j_l(\eta)} - (l+1)\right]\right\}\frac{\xi j_{l+1}(\xi)}{j_l(\xi)} - \tfrac{1}{2}\eta^4 + (l-1)(2l+1)\eta^2$$
$$+\left[\eta^2 - 2l(l-1)(l+2)\right]\frac{\eta j_{l+1}(\eta)}{j_l(\eta)} = 0, \qquad (l \neq 0) \qquad (1)$$

and

$$\frac{\tan\xi}{\xi} - \frac{4}{4-\eta^2} = 0, \qquad (l=0) \qquad (2)$$

where $j_l(\eta)$ are the spherical Bessel function of the first kind, and $\xi$ and $\eta$ are eigenvalues given by

$$\xi_{nl} = \frac{\pi v_{nl} d}{V_L} \qquad \text{and} \qquad \eta_{nl} = \frac{\pi v_{nl} d}{V_T}, \qquad (3)$$

where $V_L$ and $V_T$ are the respective longitudinal and transverse sound velocities.

These velocities were found by fitting the calculated mode frequencies, obtained from the above equations, to the measured ones based on the procedure detailed in Ref. 3. In the case of the polystyrene spheres, fitting yielded transverse and longitudinal sound velocities of 1203 and 1903 m/s, respectively. The theoretical mode frequencies, based on these parameters, were then plotted against the various sphere sizes studied in Fig. 4. The variation of the measured frequencies with sphere diameter is also presented in Fig. 4 which shows good agreement between theory and experiment. Hence, the observed Brillouin peaks were attributed to the (1,2), (1,0), (2,2) and (1,4) spheroidal modes in single polystyrene spheres.

The PMMA data were also analyzed based on the same approach. The fitted values of the transverse and longitudinal sound velocities in PMMA single spheres were found to be 1165 and 1846 m/s respectively. The acoustic mode frequencies, calculated from the sound velocities, were plotted against the sphere sizes studied. The dependence of the theoretical and experimental mode frequencies on particle size is displayed in Fig.5. As with the case of polystyrene, there is good agreement between the calculated and measured data. The observed Brillouin peaks were thus attributed to the (1,2), (1,0), (2,2), (1,4) and (1,6) spheroidal modes of a single PMMA sphere.

Based on a mass density value of $1.05 \times 10^3$ kg/m$^3$ for polystyrene, the Young's modulus and the Poisson ratio were calculated to be $3.6 \pm 0.2$ GPa and $0.17 \pm 0.01$ for the polystyrene particles. In the case of PMMA, calculations based on a density of $1.19 \times 10^3$ kg/m$^3$ yielded respective values of $3.8 \pm 0.2$ GPa and $0.17 \pm 0.01$ for the nanospheres studied. Interestingly, the Young's moduli of the nanostructures are comparable to the respective values of $3.4 \pm 0.2$ and $4.06 \pm 0.59$ GPa for bulk polystyrene and PMMA [11, 12]. In contrast, the Poisson ratios of $0.35 \pm 0.1$ for bulk polystyrene [11], and $0.34 – 0.36$ [13,14] for bulk PMMA are much larger than those of the corresponding materials in the form of nanospheres.

Dunn and Ledbetter have carried out theoretical investigations into the effect of the presence of pores in isotropic solids on their Poisson ratio [15]. They discovered that this quantity decreases with increasing pore concentration in these solids. As the smallest sphere diameter, of 300 nm, in this study is considered large for size effects to be of any real significance, it is likely that defects, such as pores, could be partly responsible for the observed reduction in the Poisson ratios relative to those of the corresponding bulk materials.

## 4. CONCLUSIONS

In summary, micro-Brillouin spectroscopy was employed in the investigation of localized acoustic modes of single isolated polystyrene and PMMA nanospheres. The study of these particles in their single isolated form by inelastic light scattering is a near realization of the ideal free-surface conditions for which Lamb's theory was formulated. The observed dependence of the confined spheroidal mode frequencies on the diameters of the nanospheres accords well with Lamb's theory. It was found that while their Young's moduli are comparable to those of the bulk materials, their Poisson ratios are much lower than those of the bulk materials. This reduction in Poisson ratios is likely due to defects, such as pores, in the nanospheres. We have demonstrated that micro-Brillouin spectroscopy is a powerful technique for probing the acoustic dynamics and mechanical properties of nanostructures.

**Acknowledgments:** Financial support from the Ministry of Education, Singapore under research grant No. R-144-000-185-112 is gratefully acknowledged.

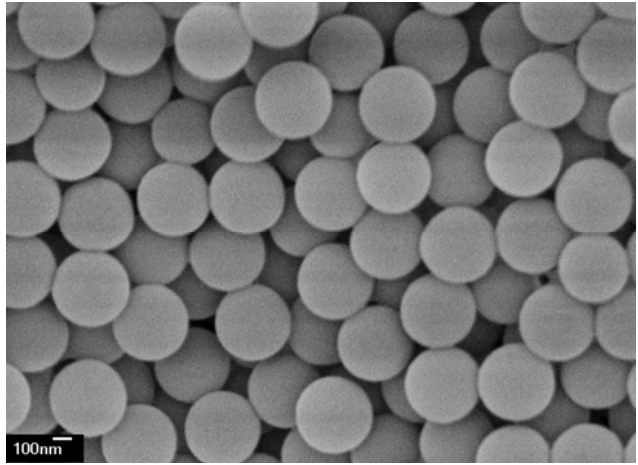

**Fig. 1.** SEM image of an aggregate of 400 nm polystyrene nanospheres.

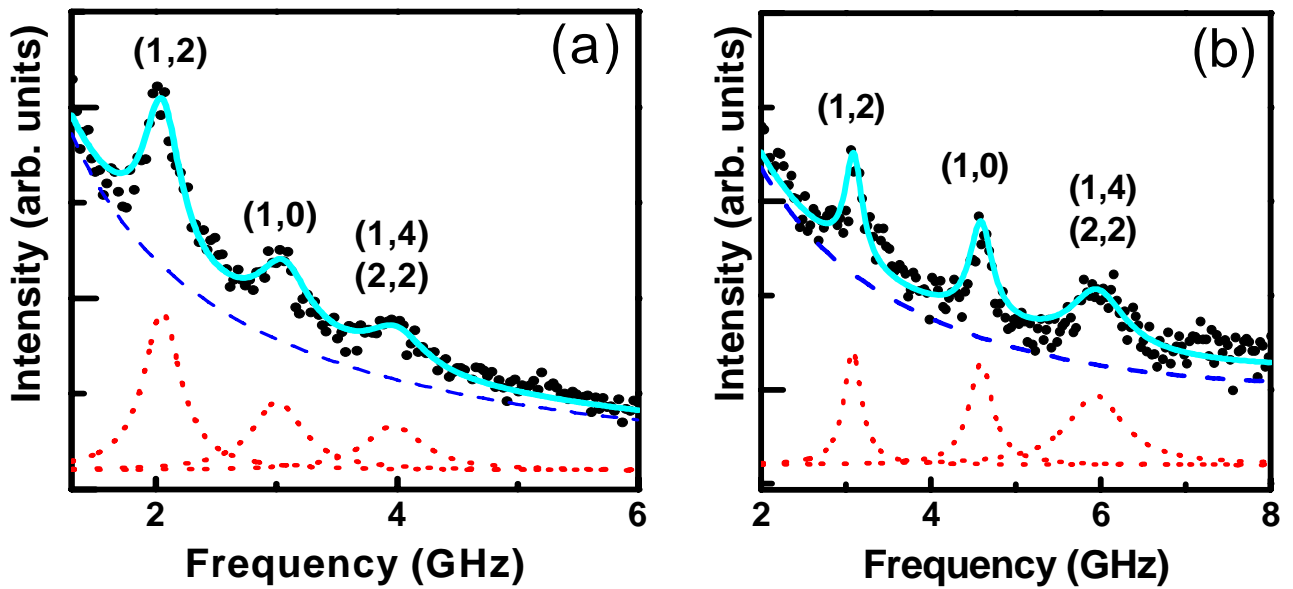

**Fig. 2.** Micro-Brillouin anti-Stokes spectra of (a) a single isolated 494 nm polystyrene sphere, and (b) a single isolated 300 nm PMMA sphere. Experimental data are denoted by dots. Each spectrum is fitted with Lorentzian functions (dotted curves) and a baseline (dashed curve), while the resultant fitted spectrum is shown as a solid curve. Confined acoustic modes are labeled by (*n,l*).

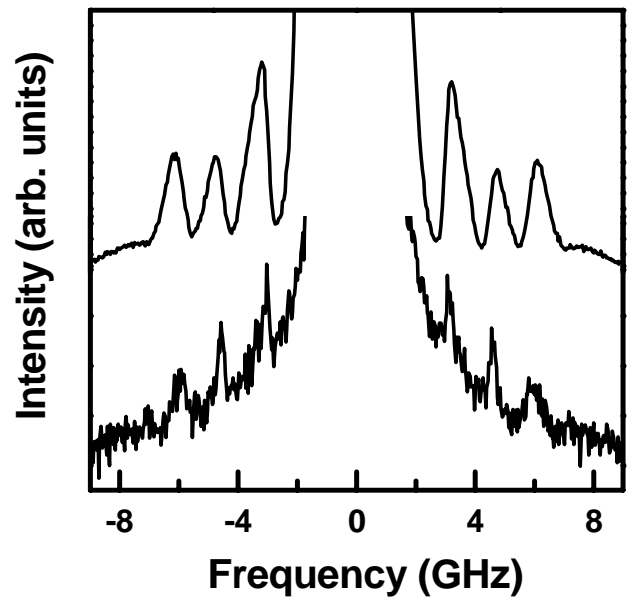

**Fig. 3.** Micro-Brillouin spectra of an aggregate of loose 300 nm PMMA spheres (top), and of a single isolated PMMA sphere (bottom).

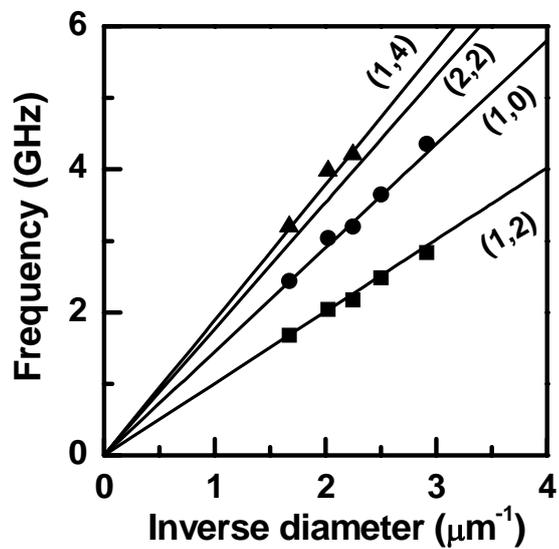

**Fig. 4**. Dependence of eigenmode frequencies of single isolated polystyrene spheres on inverse sphere diameter. Experimental data are represented by symbols for diameters $d =$ 343, 400, 445, 494 and 597 nm. The measurement errors are the size of the symbols used. The solid lines represent the theoretical frequencies of localized acoustic modes labeled by ($n,l$).

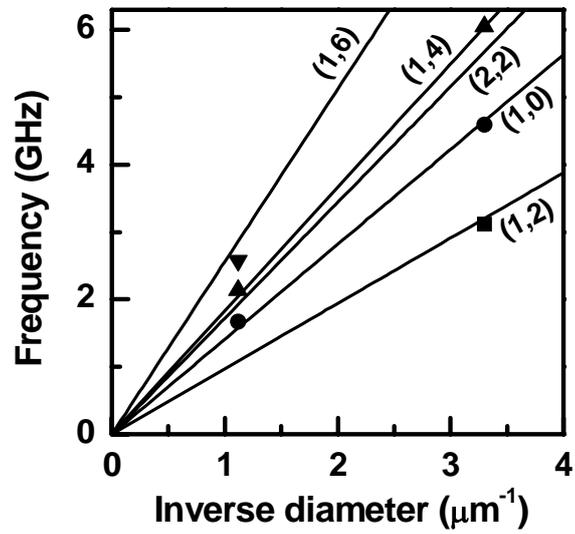

**Fig. 5.** Dependence of eigenmode frequencies of single isolated PMMA spheres on inverse sphere diameter. Experimental data are represented by symbols for diameters $d = 300$ and 900 nm. The measurement errors are the size of the symbols used. The solid lines represent the theoretical frequencies of localized acoustic modes labeled by ($n,l$).